\documentclass[11pt,a4paper]{article}
\usepackage[T1]{fontenc}
\usepackage[utf8]{inputenc}
\usepackage{lmodern}
\usepackage{amsmath,amssymb,amsthm,mathtools,mathrsfs}
\usepackage{geometry,microtype,hyperref,enumitem,booktabs,array}
\hypersetup{colorlinks=true,linkcolor=blue,citecolor=blue,urlcolor=blue}
\newcommand{\TT}{\mathbb T}
\newcommand{\RR}{\mathbb R}

\newcommand{\EE}{\mathbb E}
\newcommand{\WW}{\mathcal W}
\newcommand{\HH}{\mathcal H}
\newcommand{\dd}{\mathrm d}
\newcommand{\renH}{:\!\mathcal H\!:}

\newtheorem{theorem}{Theorem}[section]
\newtheorem{proposition}[theorem]{Proposition}
\newtheorem{remark}[theorem]{Remark}
\newtheorem{lemma}[theorem]{Lemma}
\title{Remarks on Souriau's Geometric Thermodynamics\\for Weak Two-Dimensional Euler Flows}
\author{Jean-Pierre Magnot\\[5pt]
\small Univ. Angers, CNRS, LAREMA, SFR MATHSTIC, F-49000 Angers, France\\
\small Lepage Research Institute, 17 novembra 1, 081 16 Pre\v{s}ov, Slovakia\\
\small Lyc\'ee Jeanne d'Arc, Avenue de Grande Bretagne,\\
\small 63000 Clermont-Ferrand, France\\[3pt]
\small\texttt{jean-pierr.magnot@ac-clermont.fr}}
\date{}
\begin{document}
\maketitle
\begin{abstract}
We examine infinitesimal symmetries of relative entropy on low-regularity
vorticity spaces and compare them with symmetries of stationary laws of weak
two-dimensional Euler trajectories. Relative-entropy invariance under
sample-space transformations is equivalent to preservation of the reference
law; the corresponding weak infinitesimal condition is a vanishing
reference-divergence identity. White-noise vorticity has an
infinite-dimensional family of such symmetries, generated by smooth
area-preserving transport. At the level of path laws, however, marginal
entropy symmetries need not preserve temporal correlations. We first give a
continuous Gaussian counterexample, including its infinitesimal form. We
then establish a counterexample directly for stationary white-noise weak Euler
path laws: explicit non-isometric area-preserving shears preserve the
white-noise marginal but cannot preserve an Euler path law satisfying the
standard symmetrized weak vorticity equation. We also examine the reduction
of guaranteed entropy symmetries under conditioning by renormalized energy.
These observations distinguish thermodynamic symmetry, Hamiltonian
covariance, and statistical path dynamics without asserting a selection
principle for nonunique weak solutions.
\end{abstract}



\section{Introduction}
The geometric formulation of ideal hydrodynamics originates in Arnold's
interpretation of Euler motion as geodesic dynamics on a group of
volume-preserving diffeomorphisms \cite{Arnold1966}, subsequently developed
in the analytical theory of Ebin and Marsden \cite{EbinMarsden1970}
and the geometric account of Arnold and Khesin \cite{ArnoldKhesin1998}.
In two dimensions, the vorticity formulation is naturally Lie--Poisson,
with kinetic energy as Hamiltonian and suitable vorticity functionals
as Casimirs. The smooth and bounded-vorticity regimes admit a classical
dynamical interpretation \cite{Yudovich1964}, but the passage to weak and
distribution-valued vorticities changes the available functional-analytic
structure. Diffeological approaches to approximation schemes and
infinitesimal symmetries of weak equations provide another useful
geometric viewpoint \cite{Magnot2020}.

The background in geometric thermodynamics is Souriau's theory of
Hamiltonian group actions and Gibbs ensembles \cite{Souriau,Marle2016}.
The relationship between geometric temperature, momentum maps,
entropy and information geometry has been developed by Barbaresco
\cite{Barbaresco2014,Barbaresco2016,Barbaresco2018,Barbaresco2025}
and by Barbaresco and Gay-Balmaz \cite{BarbarescoGayBalmaz2020}.
A complementary, measure-theoretic direction is Neeb's recent
classification of coadjoint orbits carrying Gibbs ensembles, where
the partition functions are integrated against Liouville measures
\cite{Neeb2026}. By contrast, the extension of Souriau's construction
to normalized means on infinite-dimensional Hamiltonian manifolds is
investigated in \cite{MagnotAoP2026,MagnotPreprint2026}. The present
article uses existing probability laws as reference normalized means;
it does not require a new reference-measure existence theorem.

The statistical mechanics of two-dimensional Euler flows is supported
by extensive results on invariant measures and generalized solutions.
Albeverio and Cruzeiro gave early constructions of invariant Gibbs
measures and associated dynamics \cite{AC1990}; Bouchet and Corvellec
studied invariant measures and equilibrium statistical mechanics
\cite{BouchetCorvellec2010}. For low-regularity vorticity, Flandoli
constructed weak Euler solutions with spatial white-noise distributions
\cite{Flandoli2018}, while Flandoli and Luo developed energy-conditional
measures and stationary statistical weak solutions
\cite{FL2020}. The distinction between stationary marginals and
probability laws on full paths is indispensable here: a law can have
time-independent one-point distributions without being invariant under
all spatial transformations that preserve those marginals.

Independently, weak Euler equations are known to admit remarkable
nonunique solutions. The pioneering examples of Scheffer
\cite{Scheffer1993} and Shnirelman \cite{Shnirelman1997} were placed
in the modern framework of differential inclusions by De Lellis and
Sz\'ekelyhidi \cite{DeLellisSzekelyhidi2009}; Villani's Bourbaki
exposition \cite{Villani2010} discusses the conceptual implications.
These works are motivational in the present note, rather than inputs
to a selection criterion: no entropy-based selection of weak solutions
is asserted here.

Our principal objective is to compare infinitesimal symmetries of
relative entropy with respect to a one-time reference measure and
symmetries of the corresponding path-space entropy. For transformations
induced on the underlying vorticity space, the former are linked to
vanishing reference divergence, whereas the latter must preserve all
finite-dimensional temporal distributions. We develop both a Gaussian
model with an explicit strict inclusion of symmetry Lie algebras and a
shear-based counterexample for weak Euler path laws. We also distinguish
symmetries preserved by renormalized-energy conditioning from general
area-preserving transformations. The resulting statements remain within
clearly stated classes of admissible transformations; they do not
classify all symmetries of the statistical state manifold.

\section{Weak vorticity path laws}
Write the two-dimensional Euler equations on $\TT^2$ as
\begin{equation}\label{eq:euler}
 \partial_t\omega+u\cdot\nabla\omega=0,\qquad u=K*\omega,
\end{equation}
where $K$ is the periodic Biot--Savart kernel.  For smooth zero-mean vorticities, a kinetic-energy Hamiltonian is
\begin{equation}\label{eq:energy}
 H(\omega)=\frac12\int_{\TT^2}|u|^2\,\dd x
 =\frac12\langle\omega,(-\Delta)^{-1}\omega\rangle.
\end{equation}
Formally, the Lie--Poisson bracket on suitable functionals is
\begin{equation}\label{eq:poisson}
 \{F,G\}(\omega)=\int_{\TT^2}\omega
 \left[\frac{\delta F}{\delta\omega},\frac{\delta G}{\delta\omega}\right]\,\dd x,
 \quad [f,g]=\partial_1f\,\partial_2g-\partial_2f\,\partial_1g.
\end{equation}
The signs in \eqref{eq:poisson} depend on the convention for the Biot--Savart operator.  None of our results below requires extending this bracket to distribution-valued paths.

Fix $\varepsilon>0$ and $T>0$, and put
\[
 E_\varepsilon=H^{-1-\varepsilon}(\TT^2),\qquad
 \WW_T=C([0,T];E_\varepsilon).
\]
The space $\WW_T$ is a Polish space with its natural topology.  For $t\in[0,T]$ denote by $e_t:\WW_T\to E_\varepsilon$ the evaluation map $e_t(\gamma)=\gamma(t)$.

A \emph{stationary-marginal weak Euler path law} is a Borel probability $P$ on $\WW_T$ which is carried by trajectories satisfying a specified weak vorticity formulation of \eqref{eq:euler} and for which
\begin{equation}\label{eq:stationary}
 (e_t)_\#P=\mu\qquad(t\in[0,T])
\end{equation}
for some Borel probability $\mu$ on $E_\varepsilon$.  Equation \eqref{eq:stationary} is not the assertion that $P$ is invariant under time shifts; on a finite time interval such an assertion in any case requires special care.

For every bounded Borel function $F$ on $\WW_T$, let
\[
 M_P(F)=\int_{\WW_T}F(\gamma)\,\dd P(\gamma).
\]
This is a positive normalized functional, and its restriction to bounded continuous or cylindrical observables is a normalized mean.  One-time marginals do not determine $P$: for instance, the expectation of
\[
 \langle\gamma(t),\varphi\rangle\langle\gamma(s),\psi\rangle
\]
may depend on the joint path law even when all one-time marginals coincide.  (For this example, the observable must be integrable; bounded cylindrical functions avoid this qualification.)

\section{Exponential reweighting and stationarity}
\begin{proposition}[Reweighting by a pathwise conserved observable]\label{prop:weight}
Let $P$ be a stationary-marginal weak Euler path law with marginal $\mu$, and let $I:E_\varepsilon\to\RR$ be Borel measurable.  Suppose that
\begin{equation}\label{eq:pathwise}
 I(e_t(\gamma))=I(e_0(\gamma))\quad P\text{-a.s. for every fixed }t\in[0,T].
\end{equation}
Let $w:\RR\to(0,\infty)$ be Borel measurable with
$0<Z_w:=\int w(I(e_0(\gamma)))\,\dd P(\gamma)<\infty$.
Then the probability law
\[
 \dd P_w(\gamma)=Z_w^{-1}w(I(e_0(\gamma)))\,\dd P(\gamma)
\]
is carried by weak Euler trajectories and has the same marginal at every time, namely
\begin{equation}\label{eq:weightedmarg}
 \dd\mu_w(x)=\frac{w(I(x))}{\int w(I)\,\dd\mu}\,\dd\mu(x).
\end{equation}
\end{proposition}
\begin{proof}
The positivity and normalization of $P_w$ are immediate.  Absolute continuity with respect to $P$ preserves the property that paths satisfy the weak equation.  For bounded Borel $f$ and fixed $t$,
\begin{align*}
 \int f(e_t(\gamma))\,\dd P_w(\gamma)
 &=Z_w^{-1}\int f(e_t(\gamma))w(I(e_0(\gamma)))\,\dd P(\gamma)\\
 &=Z_w^{-1}\int f(e_t(\gamma))w(I(e_t(\gamma)))\,\dd P(\gamma)\\
 &=Z_w^{-1}\int f(x)w(I(x))\,\dd\mu(x).
\end{align*}
This expression is independent of $t$, proving the claim.
\end{proof}

\begin{remark}[A genuine hypothesis]
The conservation property \eqref{eq:pathwise} is not a consequence of stationarity \eqref{eq:stationary}.  In particular, the energy-conditional path laws of \cite{FL2020} are shown there to have prescribed one-time marginal distributions, and their construction should not be used to assert, without an additional argument, almost-sure conservation of the renormalized energy along each path.
\end{remark}

Whenever $I$ is bounded (or appropriate exponential moments exist), take $w(z)=e^{-\theta z}$.  For
\[
 \psi(\theta)=\log\int e^{-\theta I(e_0(\gamma))}\,\dd P(\gamma),
\]
local exponential integrability justifies
\begin{equation}\label{eq:response}
 \psi'(\theta)=-\EE_{P_\theta}[I(e_0)],\qquad
 \psi''(\theta)=\operatorname{Var}_{P_\theta}(I(e_0))\ge0.
\end{equation}
These are the standard finite-parameter Gibbs response formulae expressed as path expectations.

\section{Existing energy-conditional Euler laws}
Let $\mu$ denote the Gaussian white-noise (enstrophy) measure in the framework of \cite{Flandoli2018,FL2020}.  It is supported on negative Sobolev spaces of order below $-1$.  The usual quadratic kinetic energy \eqref{eq:energy} is not a finite observable for generic white-noise vorticities.  We therefore use the renormalized-energy random variable $\renH$ constructed in \cite{FL2020}, rather than identifying it with the classical Hamiltonian at every point of $E_\varepsilon$.

For a bounded interval $I=[a,b]$ with $a<b$ the energy-conditional measure is
\begin{equation}\label{eq:conditioned}
 \mu^I(A)=\frac{\mu(A\cap\{\renH\in I\})}{p_I},
 \qquad p_I=\mu(\renH\in I)>0.
\end{equation}
Flandoli and Luo prove that there exists a weak Euler process with continuous distribution-valued trajectories and with marginal $\mu^I$ at every time \cite[Theorem~5.3]{FL2020}.  Thus, on any finite time interval, a stationary-marginal path law $P^I$ exists on a suitable $\WW_T$, without any new compactness or reference-mean construction.

The next observation combines these already constructed laws to obtain a nontrivial but elementary finite-dimensional thermodynamic family.  Its observables are \emph{labels of disjoint energy windows}; they should not be confused with the full renormalized-energy variable, nor with a pathwise conservation theorem for that variable.

\begin{proposition}[A Gibbs family from conditional Euler path laws]\label{prop:mix}
Choose pairwise disjoint bounded intervals $I_1,\dots,I_r$, with $r\ge2$, and for each $j$ let $P_j=P^{I_j}$ be a stationary-marginal weak Euler law with marginal $\mu_j=\mu^{I_j}$ supplied by \cite{FL2020}.  Let $p_j>0$ with $\sum_jp_j=1$ and choose real numbers $c_1,\dots,c_r$.  Define
\[
 P_0=\sum_{j=1}^rp_jP_j,\qquad
 Z(\theta)=\sum_{j=1}^rp_je^{-\theta c_j},\qquad
 P_\theta=Z(\theta)^{-1}\sum_{j=1}^rp_je^{-\theta c_j}P_j.
\]
Then, for every $\theta\in\RR$, $P_\theta$ is a stationary-marginal weak Euler path law with marginal
\begin{equation}\label{eq:mixtmarg}
 \mu_\theta=\sum_{j=1}^rq_j(\theta)\mu_j,
 \qquad q_j(\theta)=\frac{p_je^{-\theta c_j}}{Z(\theta)}.
\end{equation}
There exists a bounded Borel path observable $C$, equal to $c_j$ $P_j$-almost surely, such that
\begin{equation}\label{eq:gibbsmixture}
 \frac{\dd P_\theta}{\dd P_0}=\frac{e^{-\theta C}}{Z(\theta)}
 \qquad P_0\text{-a.s.}
\end{equation}
Moreover,
\begin{equation}\label{eq:mixderivatives}
 \psi'(\theta)=-\sum_jq_j(\theta)c_j,
 \qquad
 \psi''(\theta)=\sum_jq_j(\theta)\bigl(c_j-\sum_\ell q_\ell(\theta)c_\ell\bigr)^2,
\end{equation}
where $\psi=\log Z$.  The Gibbs variational identity holds:
\begin{equation}\label{eq:mixvar}
 -\log Z(\theta)=\inf_{Q\in\mathcal P(\WW_T)}
 \left\{D(Q\Vert P_0)+\theta\EE_Q[C]\right\},
\end{equation}
with the unique minimizer $P_\theta$, where $D$ denotes relative entropy and is understood to be $+\infty$ when $Q$ is not absolutely continuous with respect to $P_0$.
\end{proposition}
\begin{proof}
Finite convex combinations preserve the property of being carried by weak Euler paths.  Since every $P_j$ has stationary one-time marginals $\mu_j$, the marginal of $P_\theta$ is \eqref{eq:mixtmarg} for every $t$.

Choose a Borel version of $\renH$ on the white-noise space and set $C(\gamma)=c_j$ when $\renH(e_0(\gamma))\in I_j$, defining it arbitrarily, but boundedly, outside the union.  Because the intervals are disjoint and $P_j(e_0\in\{\renH\in I_j\})=1$, we have $C=c_j$ $P_j$-almost surely.  The measures $P_j$ are mutually singular (their time-zero supports are disjoint), and therefore \eqref{eq:gibbsmixture} follows directly from the finite mixture formula.

Differentiating the finite sum $Z(\theta)$ proves \eqref{eq:mixderivatives}.  Finally, for every $Q\ll P_0$, the relative-entropy chain rule gives
\[
 D(Q\Vert P_\theta)
 =D(Q\Vert P_0)+\theta\EE_Q[C]+\log Z(\theta).
\]
The right-hand side is minimized exactly when $Q=P_\theta$, proving \eqref{eq:mixvar}; the identity extends to arbitrary $Q$ with the usual $+\infty$ conventions.
\end{proof}

\begin{remark}[Interpretation and scope]
The potential $C$ is a coarse-grained label distinguishing energy windows.  Thus Proposition~\ref{prop:mix} is a Gibbs family on an explicitly defined finite-dimensional statistical sector of weak Euler path laws.  The resulting thermodynamic metric is the nonnegative Hessian $\psi''$, and it is strictly positive when at least two labels $c_j$ differ.  No claim is made that $C$ is an everywhere-defined classical Hamiltonian, that the renormalized energy is pathwise conserved, or that $P_\theta$ is invariant under all time shifts of a two-sided process.
\end{remark}

\section{Infinitesimal symmetries of relative entropy}\label{sec:sym}

\subsection{Entropy symmetries before dynamics}
Let $X$ be a standard Borel space with reference probability $\mu$, and
write $\mathcal P(X)$ for its probability measures.  The relative entropy is
\[
 D(\nu\Vert\mu)=\sup_{f\in\mathcal B_b(X)}
 \left\{\int_X f\,\dd\nu-\log\int_Xe^f\,\dd\mu\right\},
\]
where $\mathcal B_b(X)$ denotes the bounded Borel functions.  In the
absolutely continuous case $\dd\nu=\rho\,\dd\mu$ it equals
$\int\rho\log\rho\,\dd\mu$; it is $+\infty$ otherwise.  For this
section, an \emph{entropy symmetry induced on $X$} means a bimeasurable
bijection $T:X\to X$ whose push-forward preserves $D(\cdot\Vert\mu)$
for \emph{every} $\nu\in\mathcal P(X)$.  This requirement is stronger
than tangency to one entropy level set at one particular state.

\begin{proposition}[Exact induced entropy symmetries]\label{prop:entropysym}
For a bimeasurable bijection $T:X\to X$, the following assertions are
equivalent:
\begin{enumerate}[label=\textup{(\roman*)}]
\item $T_\#\mu=\mu$;
\item $D(T_\#\nu\Vert\mu)=D(\nu\Vert\mu)$ for every
 $\nu\in\mathcal P(X)$.
\end{enumerate}
\end{proposition}
\begin{proof}
Relative entropy is invariant under a bimeasurable change of coordinates:
\[D(T_\#\nu\Vert T_\#\mu)=D(\nu\Vert\mu),\]
as follows directly by
substituting $f\circ T$ in the variational formula.  This proves
\textup{(i)}$\Rightarrow$\textup{(ii)}.  Conversely, apply
\textup{(ii)} to $\nu=\mu$.  Since relative entropy vanishes only when
the two probability measures agree, $T_\#\mu=\mu$.
\end{proof}

Thus the entropy symmetry group induced by transformations of the sample
space is $\operatorname{Aut}(X,\mu)$.  This does not characterize
\emph{all} vector fields on an infinite-dimensional manifold of
statistical states: it characterizes those induced by changes of
coordinates on $X$.  For a density variation $\dot\rho$ with
$\int\dot\rho\,\dd\mu=0$, the formal differential
$\dd D_\rho(\dot\rho)=\int\log\rho\,\dot\rho\,\dd\mu$
is defined only under suitable positivity and integrability conditions.

\subsection{Weak infinitesimal generators}
Let $\{T_t\}_{t\in\RR}$ be a measurable one-parameter group of
$\mu$-preserving transformations of $X$.  On a stable algebra
$\mathcal C$ of cylinder observables suppose that
\[
 \mathcal L_VF=\left.\frac{\dd}{\dd t}\right|_{t=0}F\circ T_t
\]
exists in $L^1(\mu)$.  Its \emph{weak $\mu$-divergence} vanishes when
\begin{equation}\label{eq:weakdiv}
 \int_X\mathcal L_VF\,\dd\mu=0\quad(F\in\mathcal C).
\end{equation}
Invariance of $\mu$ implies \eqref{eq:weakdiv} by differentiation,
whenever this passage is justified.  Conversely, weak divergence zero
alone does not imply the existence of a global $\mu$-preserving flow;
a suitable well-posedness or core hypothesis is needed.  Under stronger
regularity assumptions the change of entropy along a continuity equation
is expressed by
\begin{equation}\label{eq:entropy-balance}
 \frac{\dd}{\dd t}D(\rho_t\mu\Vert\mu)
 =-\int_X\rho_t\,\operatorname{div}_{\mu}V\,\dd\mu,
\end{equation}
which is only asserted when the logarithmic chain rule and the
integration-by-parts operations are valid.

\subsection{Spatial white noise and area-preserving transport}
Set $X=H^{-1-\varepsilon}_0(\TT^2)$, where the zero mode is removed,
and let $\mu$ be the centered spatial white-noise law characterized by
\begin{equation}\label{eq:noise-cf}
 \int_Xe^{i\langle\omega,h\rangle}\,\dd\mu(\omega)
 =\exp\left(-\tfrac12\|h\|_{L^2}^2\right),
 \qquad h\in C^\infty_0(\TT^2).
\end{equation}
Here the subscript zero means zero spatial average.  For a smooth
orientation-preserving area-preserving diffeomorphism $g$ of $\TT^2$,
let $T_g\omega=\omega\circ g^{-1}$, interpreted on distributions by
\[
 \langle T_g\omega,h\rangle=\langle\omega,h\circ g\rangle.
\]
Since composition by $g$ is unitary on $L^2(\TT^2)$ and preserves the
mean-zero subspace, the characteristic functional \eqref{eq:noise-cf}
is unchanged.

\begin{proposition}[A geometric Lie algebra of entropy symmetries]
\label{prop:sdiff}
Every smooth area-preserving diffeomorphism $g$ acts continuously on
$X$, preserves the white-noise law $\mu$, and therefore induces an
entropy symmetry
\[
 D((T_g)_\#\nu\Vert\mu)=D(\nu\Vert\mu)
 \qquad(\nu\in\mathcal P(X)).
\]
If $v$ is a smooth divergence-free vector field on $\TT^2$ with flow
$g_t$, the induced infinitesimal action on cylinder observables
$F(\omega)=f(\langle\omega,h_1\rangle,\ldots,
\langle\omega,h_n\rangle)$ is
\begin{equation}\label{eq:infinitesimal}
 \mathcal L_vF(\omega)
 =\sum_{j=1}^n\partial_jf\,
 \langle\omega,v\cdot\nabla h_j\rangle,
 \qquad \int_X\mathcal L_vF\,\dd\mu=0
\end{equation}
for bounded smooth $f$ with bounded derivatives.  In the distributional
sense the corresponding velocity of the state is
$V_v(\omega)=-v\cdot\nabla\omega$.
\end{proposition}
\begin{proof}
The invariance follows from \eqref{eq:noise-cf}: for smooth $h$,
$\|h\circ g\|_2=\|h\|_2$.  The characteristic functional determines
the Gaussian law.  Composition by a fixed smooth diffeomorphism and its
inverse is bounded on each Sobolev distribution space $H^q$; hence
$T_g$ acts continuously on $X$.  Differentiating
$\langle T_{g_t}\omega,h_j\rangle
=\langle\omega,h_j\circ g_t\rangle$
gives \eqref{eq:infinitesimal}.  The integral of the latter vanishes
by differentiating the invariant expectation of $F\circ T_{g_t}$.
\end{proof}

\begin{remark}[Loss of one derivative]
The map $\omega\mapsto-v\cdot\nabla\omega$ generally maps
$H^{-1-\varepsilon}$ into $H^{-2-\varepsilon}$, not into the original
space.  Thus \eqref{eq:infinitesimal} is a weak infinitesimal action on
cylinder observables (or on a scale of Sobolev spaces), not an
assertion that $V_v$ is an everywhere-defined Banach-space tangent
vector field on $X$.  In particular, the algebra in
Proposition~\ref{prop:sdiff} is a concrete \emph{subalgebra} of entropy
symmetries, not a classification of every entropy-preserving state-space
transformation.
\end{remark}

A useful statistical consequence can already be computed without Euler
dynamics.  For smooth mean-zero $h_1,\ldots,h_r$, let
$J_j(\omega)=\langle\omega,h_j\rangle$.  Gaussian integration gives
\begin{equation}\label{eq:gaussian-J}
 \log\int_X\exp\Bigl(-\sum_{j=1}^r\lambda_jJ_j(\omega)\Bigr)
 \,\dd\mu(\omega)
 =\frac12\sum_{i,j=1}^r\lambda_i\lambda_j
 \langle h_i,h_j\rangle_{L^2}.
\end{equation}
These $J_j$ are well-defined weak linear observables, but are not
in general conserved by Euler evolution; \eqref{eq:gaussian-J} is not
an assertion about invariant Gibbs dynamics.

\subsection{Conditioning by renormalized energy}
Write $\mathscr E$ for a measurable version of the renormalized energy
from \cite{FL2020} and, when $\mu(\mathscr E\in I)>0$, let
$\mu^I=\mu(\,\cdot\mid\mathscr E\in I)$.  Define the subgroup
\[
 G_{\mathscr E}=\bigl\{g\in\operatorname{SDiff}(\TT^2):
 \mathscr E(T_g\omega)=\mathscr E(\omega)
 \text{ for $\mu$-a.e. $\omega$}\bigr\}.
\]

\begin{proposition}[Entropy symmetries retained after conditioning]
\label{prop:conditioning}
Every $g\in G_{\mathscr E}$ preserves $\mu^I$ for every admissible
energy window $I$.  Hence $g$ preserves the relative entropy
$D(\cdot\Vert\mu^I)$ on $\mathcal P(X)$.  In particular, the
orientation-preserving isometry group
$K=\TT^2\rtimes C_4$ of the square flat torus is contained in
$G_{\mathscr E}$, for the standard Fourier/Green-kernel
renormalized energy.
\end{proposition}
\begin{proof}
For a Borel set $A$, invariance of $\mu$ and of $\mathscr E$ under
$T_g$ gives
\[
 \mu^I(T_g^{-1}A)
 =\frac{\mu(T_g^{-1}A\cap\{\mathscr E\in I\})}
 {\mu(\mathscr E\in I)}
 =\mu^I(A).
\]
Proposition~\ref{prop:entropysym} yields the entropy claim.  The
flat isometries commute with the Laplacian and preserve the Green
kernel; the corresponding Fourier cutoff definitions of the centered
quadratic renormalized energy are isometry invariant.  Passing to the
$L^2(\mu)$ limit preserves this equality, for each fixed isometry.
\end{proof}

\begin{remark}[Scope of the conditioned symmetry statement]
The result identifies a \emph{guaranteed subgroup}, not the entire
symmetry group of a particular conditioned law.  An arbitrary
area-preserving diffeomorphism does not in general commute with the
Laplacian or preserve the kinetic-energy quadratic form.  Moreover,
the infinite-dimensional Lie algebra generated by area-preserving
transport is established for the \emph{unconditioned} white-noise
entropy, whereas the guaranteed connected spatial symmetry group
for the energy-conditioned reference is the translation torus.
Statements about a conserved Euler momentum map require further
analysis, and neither Proposition~\ref{prop:sdiff} nor
Proposition~\ref{prop:conditioning} identifies all such observables.
\end{remark}

\subsection{Comparison with Euler infinitesimal symmetries}
For a smooth vorticity the formal infinitesimal action of $v$ is
$V_v(\omega)=-v\cdot\nabla\omega$.  It is an infinitesimal symmetry of
the \emph{equation} only if the induced action commutes with the
velocity recovery $u=K*\omega$ in the corresponding formulation.
Smooth orientation-preserving isometries have this property; an
arbitrary $v$ with $\operatorname{div}v=0$ does not.  Consequently,
the area-preserving transport algebra furnishes a potentially
larger family of entropy symmetries than the spatial dynamical
symmetries of Euler.  In particular,
\begin{equation}\label{eq:strict-distinction}
 \left\{V_v:\operatorname{div}v=0\right\}
 \supseteq
 \left\{V_v:v\text{ generates a spatial Euler symmetry}\right\}.
\end{equation}
On the square flat torus, the infinitesimal spatial isometries are the
constant vector fields, generating translations; quarter-turns are
additional discrete symmetries.  These spatial translations are not
Hamiltonian vector fields generated by \emph{periodic scalar stream
functions} on $\TT^2$, since their associated closed one-forms have
nonzero periods.  Thus infinitesimal entropy symmetry, dynamical
symmetry, and existence of a Souriau momentum-map component are
three separate conditions.  In the present note only the first is
characterized in full within the class of induced bimeasurable
transformations, while concrete subgroups are compared in the
Euler setting.

\section{Entropy symmetries of weak path laws}\label{sec:path}

Let $X$ be a Polish state space and $\WW_T=C([0,T];X)$ a path space.
Only transformations $T:X\to X$ whose pointwise actions map continuous
paths to continuous paths will be considered. Such a transformation induces
$\widehat T:\WW_T\to\WW_T$ by
\begin{equation}\label{eq:pathlift}
 (\widehat T\gamma)(t)=T(\gamma(t)).
\end{equation}
Given a path law $P$, write $\mu_t=(e_t)_\#P$ for its one-time
marginals. There are now two natural relative-entropy functionals,
\[
 \HH_X(\nu)=D(\nu\Vert\mu),\qquad
 \HH_{\WW}(Q)=D(Q\Vert P),
\]
where in the first expression $\mu$ is a common stationary marginal.
The symmetries of these functionals should not be identified.

\begin{theorem}[Marginal and path entropy symmetries]\label{thm:path}
Suppose that $P$ has common marginal $\mu$ and $T:X\to X$ is a
bimeasurable bijection inducing a bimeasurable bijection $\widehat T$ on
$\WW_T$. Consider
\begin{align*}
\textup{(P)}\quad &\widehat T_\#P=P,\\
\textup{(M)}\quad &T_\#\mu=\mu,\\
\textup{(E)}\quad &D(T_\#\nu\Vert\mu)=D(\nu\Vert\mu)
                     \quad(\nu\in\mathcal P(X)).
\end{align*}
Then
\begin{equation}\label{eq:logical}
 \boxed{\textup{(P)}\Longrightarrow\textup{(M)}
 \Longleftrightarrow\textup{(E)}.}
\end{equation}
The converse $\textup{(M)}\Rightarrow\textup{(P)}$ fails, even for
continuous centered Gaussian processes on a finite-dimensional state
space. Moreover $\textup{(P)}$ is equivalent to the invariance of
$D(\cdot\Vert P)$ under $\widehat T_\#$ for all path laws.
\end{theorem}
\begin{proof}
Since $e_t\circ\widehat T=T\circ e_t$, condition (P) gives
$T_\#\mu=(e_t)_\#\widehat T_\#P=(e_t)_\#P=\mu$.
Proposition~\ref{prop:entropysym} establishes
$\textup{(M)}\Leftrightarrow\textup{(E)}$; applied on $\WW_T$ it also
proves the last assertion. A counterexample is provided below.
\end{proof}

The distinction admits a weak infinitesimal formulation. If $T_s$ is
induced by a suitable one-parameter flow and $\mathcal L_V$ and
$\widehat{\mathcal L}_V$ are its generators on cylindrical observables,
then
\begin{equation}\label{eq:weakpaths}
 \int_{\WW_T}\widehat{\mathcal L}_V F\,\dd P=0
 \quad\Longrightarrow\quad
 \int_X\mathcal L_V f\,\dd\mu=0.
\end{equation}
Indeed it suffices to take $F=f\circ e_t$. The converse generally fails;
thus, for lifted and sufficiently differentiable transformation groups,
\begin{equation}\label{eq:lie-inclusion}
 \mathfrak{sym}_{\mathrm{path}}(P)
 \subseteq \mathfrak{sym}_{\mathrm{marg}}(\mu).
\end{equation}
No identification with arbitrary vector fields on the nonlinear space of
all statistical states is implied.

\subsection{A continuous Gaussian counterexample}\label{sec:gauss-counter}
Let $X=\RR^3$, $Z\sim\mathcal N(0,I_3)$, and set
\begin{equation}\label{eq:gausspath}
 \gamma(t)=R_tZ,\qquad 0\le t\le1,\qquad
 R_t=\begin{pmatrix}
 1&0&0\\0&\cos t&-\sin t\\0&\sin t&\cos t
 \end{pmatrix}.
\end{equation}
The induced path law $P$ is concentrated on smooth paths and has
$\mu_t=\mathcal N(0,I_3)$ for every $t$. Let
$T=\operatorname{diag}(-1,-1,1)$. It is orthogonal, so it preserves
$\mu$. Yet
\[
 \EE_P[\gamma_2(0)\gamma_3(t)]=\sin t,\qquad
 \EE_{\widehat T_\#P}[\gamma_2(0)\gamma_3(t)]=-\sin t.
\]
Consequently $\widehat T_\#P\ne P$ whenever $0<t\le1$.

There is an infinitesimal version which is more appropriate to our
question. Let
\[
 B=\begin{pmatrix}0&-1&0\\1&0&0\\0&0&0\end{pmatrix},
 \qquad T_s=e^{sB}.
\]
Since $B$ is skew-symmetric, $T_s$ preserves $\mu$ for every $s$;
its generator $V(x)=Bx$ has zero Gaussian divergence. Nevertheless,
for the two-time observable $F(\gamma)=\gamma_1(0)\gamma_3(t)$,
\begin{equation}\label{eq:gaussian-derivative}
 \left.\frac{\dd}{\dd s}\right|_{s=0}
 \EE_P[F(\widehat T_s\gamma)]
 =-\sin t\ne0\qquad(0<t\le1).
\end{equation}
In fact the two Lie algebras can be computed exactly within the
orthogonal linear group. If $A$ denotes the generator of the rotations
$R_t$ in \eqref{eq:gausspath}, then
\[
 \mathfrak{sym}^{\mathrm{orth}}_{\mathrm{marg}}(\mu)
 =\mathfrak{so}(3),\qquad
 \mathfrak{sym}^{\mathrm{orth}}_{\mathrm{path}}(P)
 =\{B\in\mathfrak{so}(3):[B,A]=0\}=\RR A.
\]
Indeed the two-time covariance of $\gamma$ is
$\EE[\gamma(t)\gamma(r)^\top]=R_{t-r}$. A spatial orthogonal
transformation $U$ preserves the centered Gaussian path law if and
only if $UR_{t-r}U^\top=R_{t-r}$ for all $t,r$, equivalently
$UA=AU$. The centralizer of a nonzero generator of rotations in
$\mathfrak{so}(3)$ is its one-dimensional span. Therefore
\begin{equation}\label{eq:explicit-lie-dim}
 \dim\mathfrak{sym}^{\mathrm{orth}}_{\mathrm{marg}}(\mu)=3,
 \qquad
 \dim\mathfrak{sym}^{\mathrm{orth}}_{\mathrm{path}}(P)=1.
\end{equation}
This is a strict inclusion between explicitly computed symmetry
\emph{spaces}, not merely a comparison of two particular transformations. The
observable is unbounded but integrable with respect to the Gaussian law;
truncating it gives a bounded Borel test witnessing non-invariance as well.

\subsection{A counterexample for genuine weak Euler path laws}\label{sec:euler-counter}
We next obtain a stronger statement which uses the Euler nonlinearity,
not merely an arbitrary process with Gaussian marginals. Use coordinates
$x,y$ of period $2\pi$ on the flat torus and set
\[
 \psi=(-\Delta)^{-1}\omega,\qquad
 u_\omega=(\partial_y\psi,-\partial_x\psi),\qquad
 \mathcal B(\omega)=-u_\omega\cdot\nabla\omega.
\]
The corresponding symmetrized weak vorticity drift is denoted by
$\mathcal B_\varphi(\omega)$: on smooth vorticities it equals
$\langle\mathcal B(\omega),\varphi\rangle$, while on spatial white
noise it is the centered second-order Gaussian-chaos limit of the
symmetric Biot--Savart kernel, as in \cite{Flandoli2018,FL2020}.
For each smooth $\varphi$, this limit lies in $L^2(\mu)$.

\begin{lemma}[The quadratic shear defect and its Gaussian realization]
\label{lem:shear-chaos}
For the shear $g_s$ in \eqref{eq:shear} and every smooth test function
$\varphi$, let $H_\varphi$ denote the symmetric Biot--Savart kernel
in the weak vorticity formulation, with its usual centered
second-chaos realization. Define, away from the diagonal,
\[
 A_{s,\varphi}(x,y)
 =H_\varphi(g_sx,g_sy)-H_{\varphi\circ g_s}(x,y),
 \qquad \overline A_{s,\varphi}
 =(\Pi_0\otimes\Pi_0)A_{s,\varphi},
\]
where $\Pi_0$ is the orthogonal projection onto the mean-zero
subspace of $L^2(\mathbb T^2)$. Then $\overline A_{s,\varphi}$
is symmetric and square-integrable. With a fixed consistent choice of
Wick/second-chaos renormalization,
\begin{equation}\label{eq:chaos-representation}
 Q_{s,\varphi}:=\mathcal B_\varphi(T_s\omega)
 -\mathcal B_{\varphi\circ g_s}(\omega)
 =I_2(\overline A_{s,\varphi}),\qquad
 \EE_\mu|Q_{s,\varphi}|^2
 =2\|\overline A_{s,\varphi}\|_{L^2(\mathbb T^2\times\mathbb T^2)}^2.
\end{equation}
There exists a fixed smooth $\varphi$ for which
$\overline A_{s,\varphi}\ne0$ for all sufficiently small
nonzero $s$. Its derivative at $s=0$ is nonzero in $L^2$ as well.
\end{lemma}
\begin{proof}
The symmetrization removes the first-order singularity of the
Biot--Savart kernel: near the diagonal
$|K(x-y)|\lesssim |x-y|^{-1}$, while
$|\nabla\varphi(x)-\nabla\varphi(y)|\lesssim |x-y|$.
Thus $H_\varphi$ is bounded off the diagonal and belongs to
$L^2$. The same is true of $H_\varphi(g_sx,g_sy)$ since
$g_s$ is a smooth area-preserving diffeomorphism, and of
$H_{\varphi\circ g_s}$. The zero-mode projection is necessary
because our white noise has zero spatial mean.

The change-of-variables formula for the quadratic weak drift on
smooth mean-zero fields gives
\[
 \mathcal B_\varphi(T_s\omega)-
 \mathcal B_{\varphi\circ g_s}(\omega)
 =\iint_{\mathbb T^2\times\mathbb T^2}
 \overline A_{s,\varphi}(x,y)\omega(x)\omega(y)\,dx\,dy.
\]
On white noise, the centered (Wick) realization of this quadratic
form is $I_2(\overline A_{s,\varphi})$; the standard isometry for
a real isonormal Gaussian process gives \eqref{eq:chaos-representation}.
This equality concerns the consistently centered weak drifts,
not an undefined pointwise product of distributional vorticities.

To establish nonvanishing, take $\omega_*(x,y)=\cos x$ and
$q(x,y)=\sin x\sin y$. Direct calculation gives
$\mathcal B(\omega_*)=0$ and
\[
 D\mathcal B(\omega_*)[q]=-
 \tfrac12\sin^2x\cos y.
\]
Choose the fixed smooth test $\varphi(x,y)=\sin^2x\cos y$.
Because $T_s\omega_*=\omega_*+sq+O(s^2)$ in smooth norms,
\begin{equation}\label{eq:nonzero-shear-test}
 \iint \overline A_{s,\varphi}(x,y)
 \omega_*(x)\omega_*(y)\,dx\,dy
 =-\frac{s}{2}\int_{\mathbb T^2}
 \sin^4x\cos^2y\,dx\,dy+O(s^2),
\end{equation}
which is nonzero for all sufficiently small $s\ne0$.
The integral on the left is the deterministic quadratic form;
hence its nonvanishing proves that the mean-zero projected kernel
is nonzero in $L^2$.

Moreover, for $s$ near zero the kernels are differentiable in
$L^2$: this follows by differentiation off the diagonal and the
standard local estimates
$|D^jK(z)|\lesssim |z|^{-1-j}$, combined with
$|g_sx-g_sy|\asymp|x-y|$ and
$|\partial_s(g_sx-g_sy)|\lesssim|x-y|$.
The resulting differentiated symmetrized kernels remain locally
bounded. Differentiating \eqref{eq:nonzero-shear-test} at $s=0$
therefore also proves
$\partial_s\overline A_{s,\varphi}|_{s=0}\ne0$.
\end{proof}

\begin{theorem}[A marginal entropy symmetry which cannot lift to Euler paths]
\label{thm:euler-strict}
Let $P$ be a stationary-marginal path law of weak two-dimensional
Euler vorticity solutions with projected spatial white-noise marginal
$\mu$, in the symmetrized weak vorticity formulation of
\cite{Flandoli2018}. For sufficiently small $s\ne0$, consider the smooth
area-preserving shear
\begin{equation}\label{eq:shear}
 g_s(x,y)=(x+s\sin y,y),\qquad
 (T_s\omega)(x,y)=\omega(g_s^{-1}(x,y)).
\end{equation}
Then $T_s$ preserves $\mu$, and hence preserves relative entropy on
all one-time statistical states, but
\begin{equation}\label{eq:euler-nonlift}
 (\widehat T_s)_\#P\ne P.
\end{equation}
Indeed $ (\widehat T_s)_\#P$ cannot be carried by solutions of the
same symmetrized weak Euler equation. In addition, the generator
$V(\omega)=-\sin y\,\partial_x\omega$ has zero weak divergence with
respect to $\mu$ but fails to generate a symmetry of the Euler path law $P$.
\end{theorem}
\begin{proof}
The shear is generated by the divergence-free vector field
$v=(\sin y,0)$, so Proposition~\ref{prop:sdiff} gives
$(T_s)_\#\mu=\mu$ and vanishing marginal divergence.

We first exhibit the failure of dynamical covariance on smooth
vorticities. Take the steady Euler vorticity $\omega_*=\cos x$;
its velocity is $(0,\sin x)$ and $\mathcal B(\omega_*)=0$.
As $s\to0$,
\[
 T_s\omega_* =\cos(x-s\sin y)
 =\cos x+s\sin x\sin y+O(s^2).
\]
Put $q=\sin x\sin y$. The first variation of the Euler drift is
\begin{align}
 D\mathcal B(\omega_*)[q]
 &=-u_{\omega_*}\cdot\nabla q
   -u_q\cdot\nabla\omega_*\notag\\
 &=-\tfrac12\sin^2x\cos y\ne0,\label{eq:shear-drift}
\end{align}
where $(-\Delta)^{-1}q=q/2$. Thus for a suitable smooth test
$\varphi$ the quadratic discrepancy
\begin{equation}\label{eq:commutator}
 Q_{s,\varphi}(\omega)
 :=\mathcal B_\varphi(T_s\omega)
 -\mathcal B_{\varphi\circ g_s}(\omega)
\end{equation}
is not identically zero as a quadratic functional on smooth
mean-zero vorticities when $s\ne0$ is sufficiently small.

By Lemma~\ref{lem:shear-chaos}, with the fixed test
$\varphi=\sin^2x\cos y$, the centered discrepancy is the
second-chaos random variable $I_2(\overline A_{s,\varphi})$ and
\begin{equation}\label{eq:positive-chaos}
 \EE_\mu\bigl[|Q_{s,\varphi}|^2\bigr]
 =2\|\overline A_{s,\varphi}\|_2^2>0.
\end{equation}
This uses the mean-zero projection and the same renormalization
for both terms. It is not a statement about a pathwise product
$\omega\otimes\omega$ at arbitrary distributional vorticities.

Suppose, contrary to the claim, that $\widehat T_s\gamma$ is a weak
Euler path for $P$-almost every $\gamma$. The weak vorticity identity
for $\gamma$, tested against $\varphi\circ g_s$, and the same identity
for $\widehat T_s\gamma$, tested against $\varphi$, have the same
left-hand side. Subtraction therefore gives
\[
 \int_0^t Q_{s,\varphi}(\gamma(r))\,\dd r=0
 \quad\text{for all }t\in[0,T],\quad P\text{-a.s.}
\]
Hence $Q_{s,\varphi}(\gamma(r))=0$ for almost every $r$ on almost
every path. As $Q_{s,\varphi}\in L^1(\mu)$ and every marginal is
$\mu$, Tonelli's theorem yields
\[
 T\int_X|Q_{s,\varphi}(\omega)|\,\dd\mu(\omega)=0,
\]
contradicting \eqref{eq:positive-chaos}. Therefore transformed paths
cannot satisfy the same Euler equation almost surely, and in
particular the path law is not preserved.

For the infinitesimal assertion, Lemma~\ref{lem:shear-chaos}
gives a well-defined $L^2(\mu)$ derivative
\[
 Q'_{0,\varphi}
 =I_2\!\left(\left.\partial_s\overline A_{s,\varphi}
 \right|_{s=0}\right)\ne0.
\]
If the full one-parameter family $\widehat T_s$ preserved $P$,
then it would preserve the support of the weak Euler equation for
all sufficiently small $s$, contrary to the preceding argument.
Thus $V=-\sin y\,\partial_x\omega$ is not the generator of a
one-parameter group of $P$-preserving lifted transformations.
A stronger pointwise infinitesimal statement can be made only on a
specified residual-test domain: if $R_{\varphi,t}(\gamma)$ denotes
the weak Euler residual, then, whenever the residual has the stated
$L^1(P)$ differentiability,
\[
 \left.\frac{d}{ds}\right|_{s=0}
 R_{\varphi,t}(\widehat T_s\gamma)
 =\int_0^t Q'_{0,\varphi}(\gamma(r))\,dr.
\]
This right-hand side cannot vanish for every $t$ almost surely,
since otherwise differentiation in $t$ and stationary marginals
would force $Q'_{0,\varphi}=0$ in $L^1(\mu)$.
We do \emph{not} infer from this argument that a derivative of the
expectation of some arbitrary bounded cylindrical observable is
nonzero. Such a claim requires a separately established domain for
the infinitesimal action on path-law observables.

\end{proof}

\begin{remark}[What the Euler counterexample proves]
The theorem is independent of uniqueness or nonuniqueness of the
weak Euler process. It uses only the stationarity of its white-noise
marginal and the symmetrized weak equation. In particular, it does
not assert that a specified energy-conditioned law has exactly the
same broken symmetries, nor does it infer any anomaly of entropy
production from this non-invariance.
\end{remark}

\subsection{Energy conditioning: a second separation of symmetry spaces}
The distinction also occurs at the level of \emph{one-time} entropy
symmetries when the reference changes from $\mu$ to an
energy-conditioned law $\mu^I$. This is different from the path-law
obstruction above.

\begin{proposition}[A shear breaks some energy-conditioned entropy]
\label{prop:energybreak}
For all sufficiently small nonzero $s$, the shear $T_s$ of
\eqref{eq:shear} preserves white noise but does not preserve the
renormalized energy $\mathscr E$ of \cite{FL2020} as a random
variable modulo $\mu$-null sets. Consequently there exists a bounded
interval $I$ with $\mu(\mathscr E\in I)>0$ such that
$(T_s)_\#\mu^I\ne\mu^I$.
\end{proposition}
\begin{proof}
The energy is obtained by an $L^2(\mu)$ limit of centered quadratic
Fourier forms with kernel $(-\Delta)^{-1}$ (up to the overall sign
convention of \cite{FL2020}). If the energy were invariant under
$T_s$, the difference of the corresponding second-chaos kernels would
vanish in $L^2$. The same kernel, evaluated on smooth mean-zero
vorticities, is the difference of the usual kinetic energies
$H(T_s\omega)-H(\omega)$.

The failure of energy invariance can be checked explicitly,
without relying solely on the fact that the shear is not an
isometry. Use $x,y\in[0,2\pi)$ and normalized spatial measure
$dx\,dy/(2\pi)^2$ throughout this calculation, with
$H(\omega)=\tfrac12\langle\omega,(-\Delta)^{-1}\omega\rangle$.
For $\omega_*=\cos x$, let
\[
 \omega_s=T_s\omega_*
 =\cos x+s\sin x\sin y
 -\frac{s^2}{2}\cos x\sin^2y+O(s^3).
\]
The initial mode has $-\Delta$ eigenvalue $1$, whereas the
first-order mixed modes have eigenvalue $2$. Orthogonality yields
\begin{align*}
 H(\omega_*)&=\frac14,\\
 H(\omega_s)&=\frac14+s^2\left(-\frac18+\frac1{16}\right)
 +O(s^4)
 =\frac14-\frac{s^2}{16}+O(s^4).
\end{align*}
The $O(s^4)$ remainder follows from the evenness in $s$ of the
energy, using reflection in $y$. Hence the quadratic-form
difference is nonzero for every sufficiently small $s\ne0$.
Its symmetric mean-zero Fourier kernel must therefore be nonzero;
the same projected kernel is the second-chaos kernel of the
difference between the correspondingly renormalized random
energies (up to the overall normalization/sign convention used for
$\mathscr E$). Therefore the second-chaos kernel is
nonzero, proving
$\mathscr E\circ T_s\ne\mathscr E$ in $L^2(\mu)$.

There must then be a bounded interval $I$ with rational endpoints for
which the events $\{\mathscr E\in I\}$ and
$\{\mathscr E\circ T_s\in I\}$ differ on a set of positive
$\mu$-measure. If the conditional law $\mu^I$ were invariant, its
support event would have to be preserved modulo $\mu$-null sets,
since the densities of $\mu^I$ and of its push-forward are positive
constants on their respective support events. This contradiction
proves the claim.
\end{proof}

\subsection{Summary of the symmetry implications}
For any stationary-marginal law $P$ with marginal $\mu$, the situation
is summarized by
\begin{equation}\label{eq:diagram}
 \boxed{\widehat T_\#P=P}
 \quad\Longrightarrow\quad
 \boxed{T_\#\mu=\mu}
 \quad\Longleftrightarrow\quad
 \boxed{D(T_\#\nu\Vert\mu)=D(\nu\Vert\mu)\ \forall\nu}.
\end{equation}
Only the first implication is not reversible in general. In the
white-noise Euler setting, Theorem~\ref{thm:euler-strict} supplies a
non-reversibility example within the actual weak Euler equation.

\begin{table}[ht]
\centering\small
\renewcommand{\arraystretch}{1.23}
\begin{tabular}{@{}p{.25\linewidth}p{.30\linewidth}p{.35\linewidth}@{}}
\toprule
Level & Infinitesimal condition & Interpretation \\
\midrule
State entropy at $\mu$ & $\int_X\mathcal L_Vf\,\dd\mu=0$ & Invariance of the reference law; does not determine dynamics \\
Path entropy at $P$ & $\int_{\WW_T}\widehat{\mathcal L}_VF\,\dd P=0$ & Preservation of all joint temporal laws \\
White-noise Euler & $v$ smooth and divergence free & Marginal entropy symmetry; a shear need not be a path symmetry \\
Energy-conditioned reference & $v$ also preserves the renormalized energy & Guaranteed symmetries shrink; some shears fail already at one time \\
Euler dynamical symmetry & Commutation with the Biot--Savart drift & Distinct from both relative-entropy conditions \\
\bottomrule
\end{tabular}
\caption{Different infinitesimal symmetry requirements. The rows are not claims of equality of full symmetry algebras.}\label{tab:sym}
\end{table}

\section{Concluding remarks and outlook}
Our first observation is that the relative entropy of a reference law
has an induced symmetry group determined by preservation of that law.
For white-noise vorticity this gives an infinite-dimensional Lie
algebra of weak entropy symmetries generated by smooth area-preserving
transport. The corresponding divergence identities are understood on
cylindrical observables and do not require a Banach-valued vector
field on a negative Sobolev space.

The full path law contains additional temporal information. We have
shown that marginal entropy symmetry is necessary but not sufficient
for path entropy symmetry. Besides a simple Gaussian counterexample,
we gave an example directly for stationary white-noise weak Euler
solutions: a non-isometric area-preserving shear preserves the
marginal entropy but cannot preserve a law supported on the weak Euler
equation. We have also exhibited a related symmetry loss under
conditioning by renormalized energy. These observations compare
symmetry spaces rather than classifying every infinitesimal symmetry
of the infinite-dimensional statistical manifold.

Constructions of paradoxical weak Euler solutions involve limiting
procedures under prescribed physical constraints or parameter regimes.
The significance of such limits for the corresponding statistical
path laws---and in particular for their entropy-symmetry algebras---is
not known from the present analysis. No entropy-based selection
principle is proposed. The divergence identities and the strict
separation of marginal and path symmetries provide an initial
indication that the statistical topology of the limiting processes
may matter. Further investigations of these questions are in progress.

\end{document}